\documentstyle[preprint,aps,psfig]{revtex}

\begin{document}
\tighten

\def\gtwid{\mathrel{\raise.3ex\hbox{$>$\kern-.75em\lower1ex\hbox{$\sim$}}}}
\def\ltwid{\mathrel{\raise.3ex\hbox{$<$\kern-.75em\lower1ex\hbox{$\sim$}}}}
\def\bfig#1{\begin{figure}\centerline{\hbox{\psfig{file=#1}}}}
\def\efig{\end{figure}}
\def\H{{\cal H}}
\def\s#1{\sin\theta_#1}
\def\c#1{\cos\theta_#1}
\def\czz{\cos^2\theta_0}
\def\S_#1{\sqrt{\czz-2M/R_#1}}
\let\d=\dot
\def\Q#1{\sqrt{1-2M/R_#1}}

\preprint{NSF-ITP-97-051\qquad gr-qc/9706033}

\title{Quantum Decay of Domain Walls in Cosmology II:\\
	 Hamiltonian Approach}
\author{Shawn J. Kolitch\footnote{Electronic address:
	{\tt skolitch@pandora.physics.calpoly.edu,}
	current address: Department of Physics,
	California Polytechnic State University, San Luis Obispo, CA 93407}}
\address{Department of Physics,
	University of California,
	Santa Barbara, CA 93106-9530}
\author{Douglas M. Eardley\footnote{Electronic address:
	\tt doug@itp.ucsb.edu\hfil}}
\address{Institute for Theoretical Physics,
	University of California,
	Santa Barbara, CA 93106-4030}
\date{\today}
\maketitle
\begin{abstract}

This paper studies the decay of a large, closed domain wall in a closed
universe.  Such walls can form in the presence of a broken, discrete
symmetry.  We study a novel process of quantum decay for such a
wall, in which the vacuum fluctuates from one discrete state to another
throughout one half of the universe, so that the wall decays into pure
field energy.  Equivalently, the fluctuation can be thought of as the
nucleation of a second closed domain wall of zero size,
followed by its growth by quantum tunnelling and its collision with the
first wall, annihilating both.  We therefore study the 2-wall system
coupled to a spherically symmetric gravitational field.  We derive a
simple form of the 2-wall action, use Dirac quantization, obtain the
2-wall wave function for annihilation, find from it the barrier factor
for this quantum tunneling, and thereby get the decay probability.
This is the second paper of a series.

\end{abstract}


\section{Introduction}

It is well appreciated that domain walls can wreck a cosmological
model, and therefore it is of interest to find processes that can
destroy domain walls in the early universe.  In paper I \cite{paperI}
we introduced a novel process, the quantum decay of domain walls by global
fluctuation and quantum tunnelling.  We studied a closed universe
dominated by a single closed domain wall --- the Vilenkin-Ipser-Sikivie
or ``VIS" solution \cite{AV,IS} --- and we found an instanton that
mediates its decay into a closed universe containing pure field energy.
However some technical problems cropped up in the instanton calculation.
Therefore, in this paper we will study the same decay process by a
different technique, namely a Hamiltonian formulation and Dirac
quantization.

We must first explain why gravity is involved in this decay at all.
A domain wall in flat spacetime separates two infinite regions of
different discrete vacuum state, in the presence of a broken discrete
symmetry.  The wall cannot decay because
any quantum fluctuation into a no-wall state has an infinite barrier.
The Vilenkin solution \cite{AV} seems to describe an infinite domain wall
dressed by its gravitational field; however this spacetime is not
geodesically complete, and its complete analytic continuation \cite{IS}
can be interpreted as a closed, topologically $S^3$ universe dominated
by a closed finite $S^2$ domain wall.  This universe (we call it the
VIS solution) starts at infinite volume, collapses to a minimum
volume, at which point it halts and then re-expands to infinite volume.
The minimum radius of the domain wall is $R_{\rm min}\sim1/\sigma G$
where $\sigma$ is wall surface tension, and so gravity helps set this
scale $R_{\rm min}$\@.  This is the archetype of a universe dominated
by a domain wall, and the domain wall is classically forbidden from
collapsing to zero radius.  However, the universe is of finite volume,
$\sim R_{\rm min}^3$ near minimum, so the wall is subject to decay by global
quantum fluctuations, in which the vacuum state in one whole half of the
universe jumps to the same state as the other half.  Clearly this decay
process has a finite, albeit large, barrier factor
$\sim\sigma^{-2}G^{-3}\;$.  Thus, the domain wall decay problem becomes
a problem in quantum gravity.

The decay process can more particularly be regarded as follows.  A
second closed domain wall nucleates at zero size in the original universe,
and the two walls then approach each other by quantum tunnelling.  When
the two walls meet, they annihilate into pure field energy.
Figure 1 illustrates the 1-wall VIS universe itself, and also the
2-wall decay process.
For this reason we study the spherically
symmetric 2-wall system, coupled to a gravitational field, in this paper.
An important technical ingredient in this study is the result
of Thiemann and Kastrup \cite{KT}, who found an elegant pair
of canonical variables $(T,M)$ for spherically symmetric gravitational
field configurations.  Here we also find compatible canonical variables
for domain walls.

Quantum tunneling of domains is already well known in condensed matter
physics, and has been studied both theoretically and experimentally;
see {\it e.g,} \cite{SCB,Aw,Paul,Stamp}.  This gives hope that similar
processes can be understood in cosmology.

Section II is devoted to deriving the first main technical result of
this paper, a simple form of the effective action for the 2-wall system,
obtained by integrating out the spherically symmetric gravitational field:
\newpage
\begin{eqnarray}
&&S=\int dt \biggl\{iR_1^2\d\psi_1 + iR_2^2\d\psi_2 + M\d T	\nonumber\\
  &&\qquad\quad-\tilde N^t_1\left[\mu^2 R_1^2 - 4\Q1\sin^2\psi_1 -
	\left(1-\Q1\right)^2\right]				\nonumber\\
  &&\qquad\quad-\tilde N^t_2\left[\mu^2 R_2^2 - 4\Q2\sin^2\psi_2 -
	\left(1-\Q2\right)^2\right]\biggr\}.			\nonumber
\end{eqnarray}
Here $R_1$, $R_2$ are the radii of the two walls (defined by
Area$=4\pi R^2$), $\psi_1$ and $\psi_2$ are certain (imaginary) time
coordinates along the world sheets of the two walls, $(T,M)$ are the
Thiemann-Kastrup variables for the region between the two walls, and
$\mu\equiv 4\pi\sigma$\@.  Quantization of this action is
straightforward and is carried out in Sect.~III\@; then boundary
conditions are set, and the second main result is then derived, the
quantum tunnelling probability for domain wall decay,
\begin{equation}
	P \sim \exp\left({-2\pi/\mu^2G^3}\right).
\end{equation}
Section IV presents a slightly different quantization, in which
intrepretation is a bit clearer.  Section V discusses the results and
compares to paper I\@.  In Appendix A we review Dirac's method of
Hamiltonian quantization \cite{Dirac1,Dirac2,Dirac3}.

\bfig{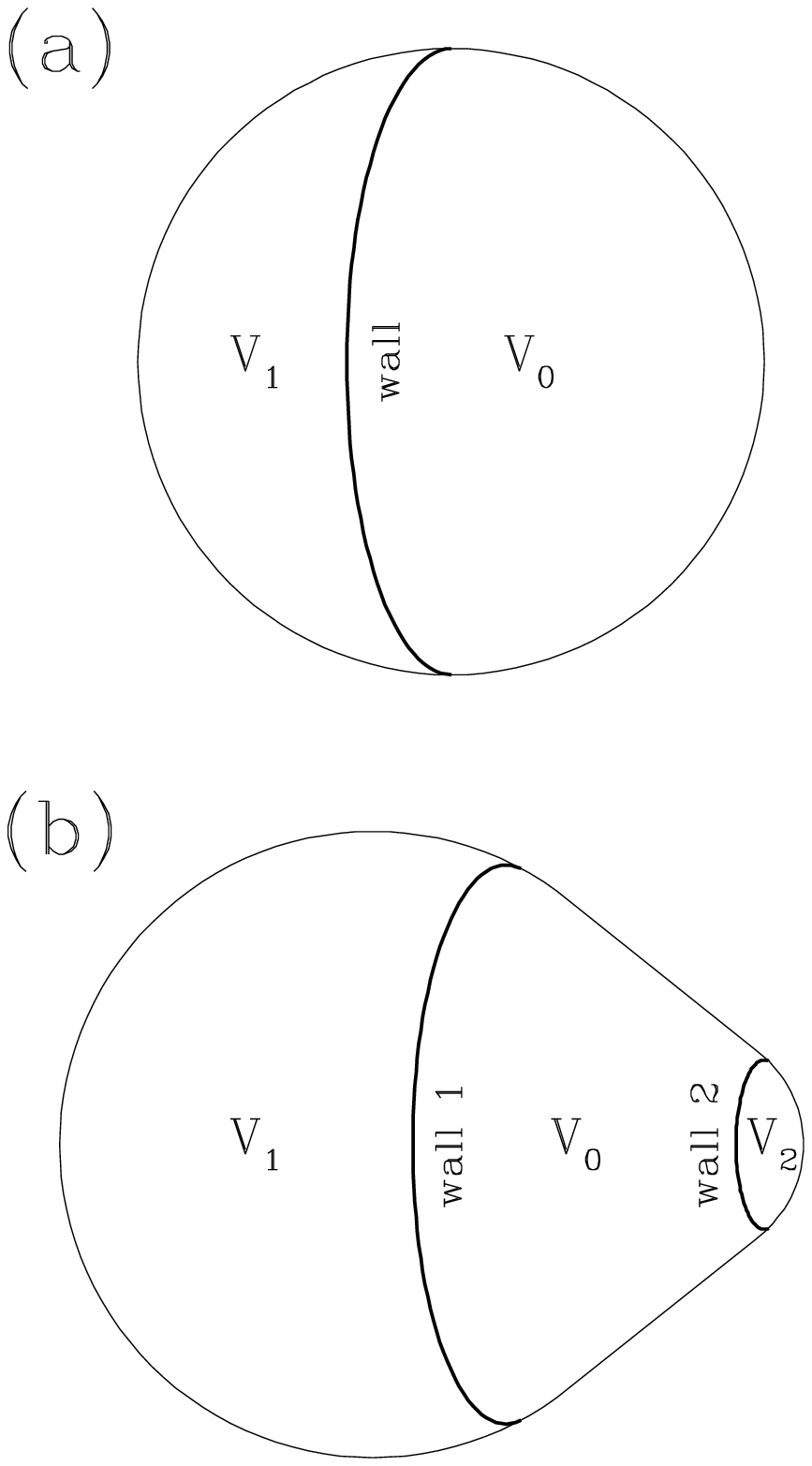,height=5.5in}
\vskip 1cm
\caption{Spaces dominated by one domain wall and by two domain walls.
The spaces are closed
and topologically $S^3$\@; heavy curves denote closed $S^2$ domain walls.
(a) A space slice of the VIS spacetime.  Regions $V_1$ and $V_0$ are
slices of flat spacetime, meeting at the domain wall.  (b) A space
slice in the quantum tunnelling regime.  The second wall has nucleated
at zero size and the two walls are tunnelling toward toward each other;
when they meet they will annihilate into pure field energy.  Regions
$V_1$, $V_0$ and $V_2$ are slices of flat Euclidean space.}
\label{q2fig1}
\efig

\section{Canonical Action for the Two Wall System}

Dirac first applied his method of Hamiltonian quantization (described
in Appendix A) to general relativity in \cite{Dirac2}, and the
theory was soon developed in greater detail by Arnowitt, Deser and
Misner \cite{ADM}.  It is not the purpose of the present study to
review the details of this subject, however, but rather to use the
canonical formalism as an alternate approach to the instanton
calculation of paper I, which was shown to have a certain pathology.
The reader is referred to the literature for a complete treatment of
canonical quantum gravity \cite{Kuchar1,Kuchar2}.

The quantization of a spherically symmetric spacetime with a finite
number of degrees of freedom is a minisuperspace model known as the
Berger-Chitre-Moncrief-Nutku (BCMN) model \cite{BCMN,Unruh,Thomi}.
We will be quantizing the spherically symmetric domain-wall spacetime
(VIS spacetime \cite{AV,IS}) introduced previously, while allowing for
the possibility that a second domain wall tunnels from zero size and
annihilates with the existing one.
\footnote{One might more generally study non-spherically-symmetric
tunnelling configurations.  We expect these to be no more probable,
and to affect the prefactor but not the exponential barrier factor in
our result.  However, we have no proof of these expectations.}
We therefore expect that there
will be just two degrees of freedom in the problem, corresponding to
the radii of the two domain walls.  A related problem with one degree
of freedom, the canonical quantization of a spherical bubble of
false-vacuum, was worked out in the WKB approximation by Fischler,
Morgan and Polchinski \cite{FMP}.  This same problem was earlier
studied using the Euclidean approach by Blau, Guendelman and Guth
\cite{BGG}, who found a pathology very similar to what was encountered
in the instanton calculation of our previous paper \cite{paperI}.

\subsection{The First Order Action}

The action for gravity plus domain walls is given in the thin-wall
limit by
\begin{equation}
	S={1\over{16\pi G}}\int d^4x\sqrt{-g}(R-2\Lambda)
		-{\mu\over{4\pi}}\int_{\hbox{walls}}d^3 A,
\end{equation}
where again $\mu/4\pi$ is $\sigma$, the energy per unit area of a
domain wall, and where
the sperically symmetric metric as
\begin{equation}
	ds^2=-(N^tdt)^2+L^2(dr+N^r)^2+R^2(d\theta^2+\sin^2\theta d\phi^2).
\end{equation}
In what follows, we will take $\Lambda=0$, corresponding to the case
of pure domain walls.  Also, except where noted, we work in units where
$G=1$.  The canonical coordinates, which are functions
of $(r,t)$, are $(N^t,N^r,L,R,r_1,r_2)$, where $r_1$ and $r_2$ are the
radial coordinates of the two domain walls.  Defining the conjugate
momenta as usual, the Hamiltonian form of this action is
\begin{equation}
	S=\int dt\left[p_1\dot r_1+p_2\dot r_2+ \int dr
		(\pi_L\dot L+\pi_R\dot R-N^t{\H}_t-N^r{\H}_r)\right],
\label{origact}
\end{equation}
where \cite{FMP}
\begin{eqnarray}
	&&{\H}_t={{L\pi_L^2}\over{2R^2}}-{{\pi_L\pi_R}\over R}
		+{1\over{2}}\left[{{2R}\over L}\left({R'\over L}\right)'
		+{R'^2\over L}-L\right]	\nonumber		\\
	&&\qquad +\sum_{j=1,2}\delta(r_j-r)\left({p_j^2\over L^2}+\mu^2
		R_j^4\right)^{1/2},\label{hdef}			\\
	&&{\H}_r=R'\pi_R-L\pi'_L-\sum_{j=1,2}\delta(r_j-r)p_j. \nonumber
\end{eqnarray}
This action is generally covariant under coordinate transformations
of $(t,r)$.

Since there are no time derivatives of $N_t$ and $N_r$ in the action,
the primary constraints are
\begin{equation}
	\pi_{N^t}=\pi_{N^r}=0,
\end{equation}
which are first-class.  However, since the Poisson brackets between
these constraints and the full Hamiltonian do not vanish, there are
the secondary constraints
\begin{equation}
	{\H}_t={\H}_r=0,
\label{hconstr}
\end{equation}
which are also first-class, and which generate coordinate transformations
of $(t,r)$\@.  Assuming $R(r)$ to be
continuous and $\pi_{L,R}$ to be free of delta-functions at each wall,
integration of these secondary constraints across a wall implies the
following jump conditions at the surface:
\begin{eqnarray}
	\Delta\pi_L&=&-{p\over L},	\nonumber		\\
	\Delta R'&=&-{{E}\over R},	\label{jump}
\end{eqnarray}
where $E\equiv(p^2+\mu^2 L^2 R^4)^{1/2}$ evaluated at the wall.

To implement these constraints in the 1-wall system, Fischler, Morgan
and Polchinski \cite{FMP} followed the Dirac approach (see the
Appendix A) to find a wave function satisfying
\begin{eqnarray}
	\pi_{N^t}\left|\Psi\right\rangle
		&=&\pi_{N^r}\left|\Psi\right\rangle=0,	\\
	{\H}_t\left|\Psi\right\rangle
		&=&{\H}_r\left|\Psi\right\rangle=0.
\label{Diracwave}
\end{eqnarray}
The first pair of these relations simply says that the wave function
is independent of the lapse and the shift; the second pair will generate
the dynamics of the wave function.

As has been mentioned, it is generally true that the first-class
constraints are in one-to-one correspondence with the gauge symmetries
of the theory; the existence of four such constraints in the present
case therefore indicates that there are four gauge degrees of freedom.
Two of these correspond to the invariance of the theory under different
choices of the lapse and shift functions; we are also free to fix the
time slicing and radial parametrization through gauge choices.  Whether
or not one fixes this part of the gauge before quantization distinguishes
Dirac quantization from ADM quantization:  Dirac's procedure, involving
no gauge fixing, leads in principle to the wave function for all
possible time slicings and radial parametrizations, whereas in the ADM
procedure one fixes the gauge before quantization and winds up with
the wave function only for a given slicing and parametrization.  The
Dirac procedure is generally more unwieldy than the ADM method; however,
one must take care that possible quantum behavior is not ruled out by
a premature gauge choice.  Indeed, it was shown in \cite{FMP} that
overzealous gauge fixing may lead to the inadvertent exclusion of parts
of the quantum dynamics.

We will therefore take a hybrid Dirac-ADM approach.  Roughly speaking,
we will ``integrate out the gravitational field":  We will fix the
radial coordinate and take a fixed family of time slices, and then solve
the constraint equations in the three vacuum regions separated by the two
walls, to reduce the action to an effective action which exclusively
involves wall degrees of freedom.  Then we will implement
Eq.~(\ref{Diracwave}) solely at the walls.

(In fact, the gauge fixing is a convenience but not a necessity for
this problem.  A future paper in this series will present a [nearly]
gauge invariant derivation of the effective action for $n$ walls in
spherical symmetry.)

\subsection{Gauge Fixing}

We first look for a solution to the constraints as follows.  Let the
radial coordinate take the range $0\le r\le r_3$, and divide the
(compact) space into three regions.  There are two centers
of spherical symmetry, located at $r=0$ and $r=r_3$, corresponding
to the centers of the two spherical domain walls, and in addition
there is a middle region between the two walls, where $r_1\le r\le r_2$.
The walls themselves are located at $r=r_1$ and $r=r_2$.  We will
refer to the three regions as region $V_1$, $V_0$, $V_2$ respectively:
\begin{eqnarray}
	V_1&:&\quad 0\le r\le r_1,	\nonumber\\
	V_0&:&\quad r_1\le r\le r_2,	\\
	V_2&:&\quad r_2\le r\le r_3.	\nonumber
\end{eqnarray}
We now fix the radial parametrization everywhere by imposing the
coordinate gauge condition
\begin{equation}
	L=1,
\end{equation}
and then impose the slicing condition
\begin{equation}
	\cases{R\pi_R=2\pi_L,	&$V_1$,\cr
		R\pi_R=\pi_L,	&$V_0$,\cr
		R\pi_R=2\pi_L,	&$V_2.$}
\end{equation}

\subsection{Solution of the Constraints}

Combining these conditions with the spatial constraint equation
\begin{equation}
	{\H}_r=R'\pi_R-\pi_L
\end{equation}
leads to the solutions
\begin{eqnarray}
	\pi_R&=&\cases{ 2ik_1 R,	&$V_1$,\cr
			ic_0,		&$V_0$,\cr
			2ik_2 R,	&$V_2$,\cr}	\nonumber\\
	\pi_L&=&\cases{ik_1 R^2,	&$V_1$,\cr
			ic_0 R,		&$V_0$,\cr
			ik_2 R^2,	&$V_2$;\cr}
\label{pi(R)}
\end{eqnarray}
where $k_1$, $c_0$, $k_2$ are constants of integration.  Here and below,
the constants of integration that appear will become our degrees of
freedom, and should all be understood as functions of time in the
dynamical problem.   The factors of $i$ have been chosen appropriate
to the classically forbidden, tunnelling regime.

The Hamiltonian constraint ${\H}_t=0$ then becomes
\begin{equation}
	0=2{\H}_t=\cases{3k_1^2 R^2-1+2RR''+R'^2,	&$V_1$,\cr
			c_0^2-1+2RR''+R'^2,		&$V_0$,\cr
			3k_2^2 R^2-1+2RR''+R'^2,	&$V_2$.}
\end{equation}
These equations have the general solutions
\begin{equation}
	r=\int{{dR}\over\sqrt{1-k_1^2 R^2}}
\end{equation}
in region $V_1$ (assuming regularity at the origin $r=0$);
\begin{equation}
	r-r_0=\int{{dR}\over\sqrt{1-c_0^2-2M/R}}
\end{equation}
in region $V_0$, where $M$ and $r_0$ are further constants of integration;
and
\begin{equation}
	r - r_3 =\int{{dR}\over\sqrt{1-k_2^2 R^2}}
\end{equation}
in region $V_3$ (assuming regularity at the anti-origin $r=r_3$)
where $r_3$ is another constant of integration.

In region $V_1$, requiring $R(0)=0$ gives the solution
\begin{equation}
	R(r)={1\over{k_1}}\sin(k_1 r),\quad 0\le r\le r_1.
\label{RI}
\end{equation}
Similarly, in region $V_2$, requiring $R(r_3)=0$ gives the solution
\begin{equation}
	R(r)={1\over{k_2}}\sin[k_2(r_3-r)],\quad r_2\le r\le r_3.
\label{RIII}
\end{equation}
In region $V_0$, defining
\begin{equation}
	c_0\equiv\sin\theta_0
\end{equation}
we leave the solution in implicit form as
\begin{equation}
	r-r_0=\int{{dR}\over\sqrt{\cos^2\theta_0-2M/R}},
		\quad r_1\le r\le r_2.
\label{RII}
\end{equation}
Here the constants $r_0$ and $r_3$ are fixed in terms of the other
variables by our requirement that $R(r)$ be continuous across each wall.
Eqs.~(\ref{RI})--(\ref{RII}) represent a general solution to the
Hamiltonian and spatial constraints, parametrized by
$(\theta_0,k_1,k_2,r_1,r_2,M)$.  It will be convenient in what follows
to define
\begin{eqnarray}
	R_j &\equiv& R(r_j) \quad(j=1,2),		\nonumber\\
	\theta_1 &\equiv& k_1r_1,			\\
	\theta_2 &\equiv& \pi-k_2(r_3-r_2);		\nonumber
\end{eqnarray}
here and throughout, the index $j=1,2$ runs over the two walls.  We
take as our independent parameters the set
$(\theta_0,\theta_1,\theta_2,M,r_1,r_2)$.

Given the above solution to the constraints, it follows from
Eq.~(\ref{pi(R)}) that
\begin{equation}
	\pi_L=\cases{
		iR\sin(k_1 r),		&$V_1$,\cr
		iR\sin\theta_0,		&$V_0$,\cr
		iR\sin[k_2(r_3-r)],	&$V_2$,\cr}
\label{piL}
\end{equation}
and
\begin{equation}
	\pi_R=\cases{
		2i\sin(k_1 r),		&$V_1$,\cr
		i\sin\theta_0,		&$V_0$,\cr
		2i\sin[k_2(r_3-r)],	&$V_2$.\cr}
\label{piR}
\end{equation}

\subsection{Reduction of the Action}

We have now found a 6-parameter family of solutions which identically
solve the constraints ${\H}_{t,r}=0$ everywhere except at the 2 walls,
while at the 2 walls we still have 2 canonical momenta $p_j$, and 4
constraints, the jump conditions (\ref{jump}).  Our reduced phase space
is now of finite dimension 8, with coordinates
$(\theta_0,\theta_1,\theta_2,r_1,r_2,M,p_1,p_2)$.  Our system is
described as a time dependent point in phase space obeying the
constraints.

We can therefore write the total action as
\begin{eqnarray}
	S=\int dt\biggl\{p_1\dot r_1+p_2\dot r_2+ \int dr
		(\pi_L\dot L+\pi_R\dot R-N^t{\H}_t-N^r{\H}_r)\biggr\}
							\nonumber\\
	 =\int dt\biggl\{\sum_{j=1,2}\left(p_j\dot r_j-N^t_j\left
		[E_j+R_j(\Delta R')_j\right]-N^r_j\left[p_j+
		(\Delta\pi_L)_j\right]\right)\nonumber\\
		+\int dr(\pi_L\dot L+\pi_R\dot R)\biggr\}.
\label{totact}
\end{eqnarray}
where $E_1$ and $E_2$, obtained from the definition
$E_j\equiv(p_j^2+\mu^2 R_j^4)^{1/2}$, are now given by
\begin{eqnarray}
	&&E_1=\left[\mu^2 R_1^4-R_1^2
		(\sin\theta_1-\sin\theta_0)^2\right]^{1/2},\nonumber\\
	&&E_2=\left[\mu^2 R_2^4-R_2^2
		(\sin\theta_0-\sin\theta_2)^2\right]^{1/2};
\label{E}
\end{eqnarray}
and where, using Eqs.~(\ref{RI},\ref{RIII},\ref{RI},\ref{piL}),
\begin{eqnarray}
	(\Delta R')_1  &=&\cos\theta_1-\S_1,		\nonumber\\
	(\Delta\pi_L)_1&=&-iR_1(\sin\theta_1-\sin\theta_0),\nonumber\\
	(\Delta R')_2  &=&-\cos\theta_2+\S_2,		\nonumber\\
	(\Delta\pi_L)_2&=&-iR_2(\sin\theta_0-\sin\theta_2).
\end{eqnarray}

Next we calculate the gravitational contribution to the action, {\it i.e.,}
\begin{equation}
	S_G = \int dtdr(\pi_L\dot L+\pi_R\dot R).
\end{equation}
In the gauge $L=1$, we can neglect the first term.  The second term is
calculated as follows:  write Eqs.~(\ref{RI})--(\ref{RII}) in the
form
\begin{equation}
	R(r) = A\Theta_A + B\Theta_B + C\Theta_C,
\end{equation}
where $A$, $B$, $C$ are just different names for $R(r)$ in the three regions,
and
\begin{eqnarray}
	&&\Theta_A = \Theta(r_1-r),	\nonumber\\
	&&\Theta_B = \Theta(r-r_1)-\Theta(r-r_2),\\
	&&\Theta_C = \Theta(r-r_2)	\nonumber
\end{eqnarray}
are step functions for the three regions.  Similarly, write
Eq.~(\ref{piR}) as
\begin{equation}
	\pi_R = 2ik_1 A\Theta_A +i\sin\theta_0\Theta_B
		+2ik_2 C\Theta_C.
\end{equation}
Then it follows that
\begin{equation}
	\pi_R\dot R=ik_1 (A^2)^{\bullet}\Theta_A+i\sin\theta_0\dot B\Theta_B
		+ik_2 (C^2)^{\bullet}\Theta_C.
\end{equation}
Integrating by parts and using the fact that
\begin{eqnarray}
	&&\dot\Theta_A=\dot r_1\delta(r_1-r)	\nonumber\\
	&&\dot\Theta_B=-\dot r_1\delta(r-r_1)+\dot r_2\delta(r-r_2)\\
	&&\dot\Theta_C=-\dot r_2\delta(r-r_2),	\nonumber
\end{eqnarray}
one finds
\begin{eqnarray}
	S_G &=&\int dt\biggl\{\dot r_1(-ik_1R_1^2+i\sin\theta_0 R_1)
		+\dot r_2(ik_2R_2^2-i\sin\theta_0 R_2)\nonumber\\
		&&\qquad\quad+\int dr(-i\dot k_1^2 A^2\Theta_A
		-i\cos\theta_0\dot\theta_0 B\Theta_B
		-i\dot k_2 C^2\Theta_C)\biggr\}	\label{gravact}\\
	&=&\int dt\biggl\{(\Delta\pi_L)_1\dot r_1+(\Delta\pi_L)_2\dot r_2 +
		\int dr(-i\dot k_1 A^2\Theta_A
		-i\cos\theta_0\dot\theta_0 B\Theta_B
		-i\dot k_2 C^2\Theta_C)\biggr\}.	\nonumber	
\end{eqnarray}
where $\theta_1\equiv k_1r_1, \; \theta_2\equiv \pi-k_2(r_3-r_2)$.
Combining Eq.~(\ref{gravact}) with Eq.~(\ref{totact}), the action
becomes
\begin{eqnarray}
	S=\int dtdr(-i\dot k_1 A^2\Theta_A
		-i\cos\theta_0\dot\theta_0 B\Theta_B
		-i\dot k_2 C^2\Theta_C)		\nonumber\\
		+\int dt\sum_{j=1,2}\left\{-N^t_j\left[E_j+R_j
		(\Delta R')_j\right] + \left[\dot r_j - N^r_j\right]
		\tilde p_j\right\},
\label{thisact}
\end{eqnarray}
where

\begin{eqnarray}
	\tilde p_1 &\equiv p_1 + (\Delta\pi_L)_1
	&\equiv p_1-iR_1(\sin\theta_1-\sin\theta_0),	\nonumber\\
	\tilde p_2 &\equiv p_2 + (\Delta\pi_L)_2
	&\equiv p_2-iR_2(\sin\theta_0-\sin\theta_2).
\end{eqnarray}

Now consider the terms involving $A$ and $C$ in Eq.~(\ref{thisact})\@.
Performing the radial integration leads to
\begin{eqnarray}
	&&\int dtdr(-i\dot k_1^2 A^2\Theta_A-i\dot k_2 C^2\Theta_C)
						\nonumber\\		
		&&=\int dt\left[-{{i\dot k_1}\over{2k_1^2}}
		(r_1-R_1\cos\theta_1)-{{i\dot k_2}\over{2k_2^2}}
		(r_3-r_2+R_2\cos\theta_2)\right].
\end{eqnarray}
Furthermore, using the definitions $k_1 R_1=\sin\theta_1, \;\theta_1=k_1
r_1$ and $k_2 R_2=\sin\theta_2, \; \theta_2=\pi-k_2(r_3-r_2)$, one
can show that
\begin{eqnarray}
	&&{{\dot k_1}\over k_1^2}(r_1-R_1\cos\theta_1)=
		R_1^2\dot\theta_1+{{du}\over{dt}},	\nonumber\\
	&&{{\dot k_2}\over k_2^2}(r_3-r_2+R_2\cos\theta_2)=
		-R_2^2\dot\theta_2+{{dv}\over{dt}},
\end{eqnarray}
where
\begin{eqnarray}
	&&u=-{R_1^2\over{2\sin^2\theta_1}}
		(\theta_1-\sin\theta_1\cos\theta_1),	\nonumber\\
	&&v=-{R_2^2\over{2\sin^2\theta_2}}
		(\pi-\theta_2+\sin\theta_2\cos\theta_2).
\end{eqnarray}
Hence we can write, dropping the total time derivatives,
\begin{equation}
	\int dtdr(-i\dot k_1^2 A^2\Theta_A-i\dot k_2 C^2\Theta_C)
		=\int dt(-iR_1^2\dot\theta_1/2+iR_2^2\dot\theta_2/2).
\label{AC}
\end{equation}
The term involving $B$ can be rewritten as
\begin{eqnarray}
	&&{}\int dtdr(-i\cos\theta_0\dot\theta_0 B\Theta_B)=
		\int dt(-i\cos\theta_0\dot\theta_0)\int_1^2
		{{RdR}\over R'}		\nonumber\\
	&&=\int dt(-i\dot\theta_0)\left[F(M,R_1,R_2)+
		{R_2^2\over2}-{R_1^2\over2}\right],
\label{Bterm}
\end{eqnarray}
where
\begin{equation}
	F(M,R_1,R_2)\equiv\int_1^2 RdR\left[{\cos\theta_0\over
		{\sqrt{\cos^2\theta_0-2M/R}}}-1\right].
\label{Fdef}
\end{equation}
Combining Eqs.~(\ref{thisact}), (\ref{AC}) and (\ref{Bterm}),
we find the reduced action
\begin{eqnarray}
	S&=&\int dt\biggl\{iR_1^2\dot\theta_1/2 + iR_2^2\dot\theta_2/2
	 - i\dot\theta_0\left(
	R_2^2/2-R_1^2/2+F(M,R_1,R_2)\right)	\nonumber\\
	&&\quad	-\sum_{j=1,2}\left\{N^t_j\left[E_j+R_j(\Delta R')_j\right]+
		\left[\dot r_j - N^r_j\right]\tilde p_j\right\}
\biggr\}.
\label{intact0}
\end{eqnarray}
Consider the last terms inside the sum over walls, the terms
proportional to $\tilde p_j$, in Eq.~(\ref{intact0})\@. From the
spatial constraints, $\tilde p_j = 0$ for all time, so these terms
make no contribution to the equations of motion, and can be dropped
from the action.  The old, gauge-dependent canonical wall coordinates
$r_j$ therefore disappear from the action, in favor of the gauge
invariant wall quantities $R_j\equiv R(r_j)$\@.  Moreover, the old,
gauge-dependent wall canonical momenta $p_j$ have completely decoupled
from the remainder of the action, of their own accord, and can be
dropped henceforth, along with the spatial constraints.
The quantities $i\theta_1$, $i\theta_2$ now act as
canonical coordinates for the two walls, and the quantities
$R_j^2/2\equiv R^2(r_j)/2$ now act as gauge-invariant canonically
conjugate momenta for the two walls.  The quantity $i\theta_0$ is a
single remaining canonical coordinate of the gravitational field in
region $V_0$, and $M$ is some kind of momentum belonging to it, though
not canonically conjugate.  Evidently, spherically symmetric gravity is
one of those simple gauge theories wherein the unphysical degrees of
freedom decouple of their own accord when appropriate canonical
coordinates are chosen.

The Hamiltonian constraints at the walls can also be simplified.
They now read
\begin{eqnarray}
	&&\sum_{j=1,2}\biggl\{-N^t_j\left[E_j+R_j
		(\Delta R')_j\right]\biggr\}=		\\
	&&-N^t_1\biggl\{\left[\mu^2 R_1^4-R_1^2
		(\sin\theta_1-\sin\theta_0)^2\right]^{1/2}
		+R_1(\sqrt{\cos^2\theta_0-2M/R_1}
		-\cos\theta_1)\biggr\}			\nonumber\\
	&&-N^t_2\biggl\{\left[\mu^2 R_2^4-R_2^2
		(\sin\theta_0-\sin\theta_2)^2\right]^{1/2}
		+R_2(\cos\theta_2 -
		\sqrt{\cos^2\theta_0-2M/R_2}\;)\biggr\}. 
\label{explconstr}
\end{eqnarray}
We have here the usual awkwardness that the $E_j$ from Eqs.~(\ref{E})
contain a square root, and in order to obtain a simpler quantum
mechanical system, we take the usual remedy and ``square out" these
constraints as follows.  Defining
\begin{equation}
N^t_j \equiv R_j^{-2}\tilde N^t_j\left[E_j-R_j(\Delta R')_j
				\right]\quad (j=1,2)
\label{tildeN}
\end{equation}
the constraints become
\begin{eqnarray}
	&&\sum_{j=1,2}\biggl\{-\tilde N^t_j\left[E_j^2/R_j^2
		(\Delta R')_j^2\right]\biggr\}=		\nonumber\\
	&&-\tilde N^t_1\left[\mu^2 R_1^2 - 1 +
	2(\c1\S_0 + \s1\s0) -1+2M/R_1\right]		\nonumber\\
	&&-\tilde N^t_2\left[\mu^2 R_2^2 - 1 +
	2(\c2\S_0 + \s2\s0) -1+2M/R_2\right].
\label{explconstr2}
\end{eqnarray}
The extra factors introduced into the constraint by Eq.~(\ref{tildeN})
never vanish in the classical regime, and make no difference to the
classical equations of motion.  Moreover they never vanish in the
quantum tunnelling regime and so at most affect the prefactor in the
tunnelling calculation.  The effective action is now
\begin{eqnarray}
	S&=&\int dt\biggl\{iR_1^2\dot\theta_1/2 + iR_2^2\dot\theta_2/2
	 - i\dot\theta_0\left(
		R_2^2/2--R_1^2/2+F(M,R_1,R_2)\right)	\nonumber\\
	&&-\tilde N^t_1\left[\mu^2 R_1^2 - 1 +
		2(\c1\S_0 + \s1\s0) -1+2M/R_1\right]		\nonumber\\
	&&-\tilde N^t_2\left[\mu^2 R_2^2 - 1 +
		2(\c2\S_0 + \s2\s0) -1+2M/R_2\right]\biggr\}.
\label{intact}
\end{eqnarray}
and the reduced phase space is now 6-dimensional with coordinates
$(\theta_0,\theta_1,\theta_2,M,R_1,R_2)$\@.

\subsection{A Cyclic Time Coordinate}

We now wish to find a further coordinate transformation in phase
space to canonical form.
In Eq.~(\ref{intact}), the quantity $M$ which appears in the integral
expression for $F(R_1,R_2)$ is the Schwarzschild mass of the region of
spacetime between the two walls.  Hence we expect, and confirm, that
variations of this action lead to $\dot M=0$ as the equation of
motion for $M$.  However, it is clear that there is no cyclic coordinate
present for which $M$ is the conjugate momentum.  $M$ is {\it part} of
the conjugate momentum of $\theta_0$, but $\theta_0$ is manifestly not
cyclic in the action.  It is thus desirable to find the coordinate
to which $M$ is canonically conjugate, and which would therefore be
cyclic in the action.

Recently, Thiemann and Kastrup found \cite{KT} that such a canonically
conjugate pair of observables can always be found for spherically
symmetric field configurations; see also \cite{KT2,Karel,Don}.
They worked in the Ashtekar approach
to canonical quantum gravity, but their result is general.  They showed
that for the line element
\begin{equation}
	ds^2=-(N^tdt)^2+L^2(dr+N^r)^2+R^2(d\theta^2+\sin^2\theta d\phi^2),
\end{equation}
the time variable conjugate to the Schwarzschild mass is given by
\begin{equation}
	T=-\int dr(1-2M/R)^{-1}\sqrt{(dR/dr)^2-L^2(1-2M/R)},
\end{equation}
which in our setting is given by
\begin{equation}
	T = -i\s0\int_{R_1}^{R_2}{dR\over\S_{{\null}}(1-2M/R)}.
\end{equation}
We would like to relate this to the function $F$ defined by
Eq.~(\ref{Fdef}).  To do so, note that
\begin{eqnarray}
	&&{{\partial T}\over{\partial R}}_{1,2}=
		-i\sin\theta_0\left[{1\over{(1-{{2M}\over R_j})
		(\cos^2\theta_0-{{2M}\over R_j})^{1/2}}}
		\right]^{R_2}_{R_1},
							\nonumber\\
	&&{{\partial T}\over{\partial\theta_0}}=
		-i\cos\theta_0\int{{dR}\over
		(\cos^2\theta_0-{{2M}\over R})^{3/2}},
\label{Tpartials}						 \\
	&&{{\partial T}\over{\partial M}}=
		-i\sin\theta_0\biggl[\int{{dR/R}\over
		{(1-{{2M}\over R})(\cos^2\theta_0-{{2M}\over R})^{3/2}}}
							\nonumber\\
		&&\qquad\qquad+2\int{{dR/R}\over{(1-{{2M}\over R})^2
		(\cos^2\theta_0-{{2M}\over R})^{1/2}}}\biggr].
							\nonumber
\end{eqnarray}
We can integrate by parts in the first term in $\partial T/\partial M$,
after writing the integrand as\hfil\break
$[R/(1-2M/R)]\cdot[dR/[R^2(\cos^2\theta_0-2M/R)^{3/2}]$ to give

\begin{equation}
	{{\partial T}\over{\partial M}}=
		\left[{{i\sin\theta_0 R_j}\over{M(1-{{2M}\over R_j})
		(\cos^2\theta_0-{{2M}\over R_j})^{1/2}}}
		\right]^{R_2}_{R_1}+{T\over M}.
\end{equation}
The time derivative of $T$ is then given by
\begin{equation}
	\dot T={{\partial T}\over{\partial R_1}}\dot R_1
		+{{\partial T}\over{\partial R_2}}\dot R_2
		+{{\partial T}\over{\partial\theta_0}}\dot\theta_0
		+{{\partial T}\over{\partial M}}\dot M.
\end{equation}
Furthermore, we can integrate $F$ by parts to find
\begin{eqnarray}
	-i\dot\theta_0 F&=&-i\dot\theta_0\cos\theta_0
		\int{{RdR}\over(\cos^2\theta_0-{{2M}\over R})^{1/2}}
		+\left[{i\over2}R^2_j\dot\theta_0\right]^{R_2}_{R_1}
						\nonumber\\
	&=&-{{i\dot\theta_0\cos\theta_0 M}\over2}
		\int{{dR}\over(\cos^2\theta_0-{{2M}\over R})^{3/2}}
						\nonumber\\
		&&-\left[{{i\dot\theta_0\cos\theta_0 R_j^2}\over
		{2(\cos^2\theta_0-{{2M}\over R_j})}
		^{1/2}}	\right]^{R_2}_{R_1}
		+\left[{i\over2}R^2_j\dot\theta_0\right]^{R_2}_{R_1}.
\label{Fbyparts}
\end{eqnarray}
Combining Eqs.~(\ref{Tpartials})--(\ref{Fbyparts}), we find that $T$
and $F$ are related by
\begin{eqnarray}
  &&{1\over2}(M\d T - \d M T) = -i\d\theta_0 F(R_1,R_2,\c0,M) - \left[
	{i\over2}R_j^2\d\theta_0\right]_{R_1}^{R_2}\nonumber\\
	&&+\left[{i\over2\S_j}\left(
	 R_j^2\c0\d\theta_0 + \s0{\d MR_j - M\d R_j\over1-2M/R_j}
	\right)\right]_{R_1}^{R_2}.
\label{TFrel}
\end{eqnarray}
Up to a total time derivative, therefore, we just get the desired
term $M\d T$ in the action, plus some messy terms at the two walls.

\subsection{Canonical Wall Coordinates}

Our next task is to find compatible canonical coordinates $(\psi_j,R_j)$
for each wall.  Consider the $N^t_1$ constraint at wall 1\@.
The angles $\theta_0$ and $\theta_1$ both enter into this constraint,
but we would like to reduce this to a dependence upon a single angle
$\psi_1$\@.  Guessing the answer, we can do this by defining $\psi_1$
by
\begin{equation}
	\sin(\theta_1+2\psi_1)	= \sin\theta_0/\Q1
\end{equation}
which implies
\begin{equation}
	\cos(\theta_1+2\psi_1)	= \S_1/\Q1
\end{equation}
Then the constraint becomes
\begin{equation}
	R_1^2 = 4\Q1\sin^2\psi_1 + (1-\Q1)^2,
\end{equation}
$\theta_0$ and $\theta_1$ themselves having disappeared as desired.
Also, by time-differentiating the definition of $\psi_1$ we find
\begin{equation}
	\d\theta_1+2\d\psi_1 = {1\over\S_1}\left(
	\c0\d\theta_0 +  \s0{\d MR_1 - M\d R_1\over R_1^2(1-2M/R_1)}
	\right)
\end{equation}
an expression which furnishes exactly the wall terms that are needed
in Eq.~(\ref{TFrel})for $\dot T${\null}.

Similar equations hold at wall 2{\null}:
\begin{equation}
	\sin(\theta_2-2\psi_2)	= \sin\theta_0/\Q2,
\end{equation}
which implies
\begin{equation}
	\cos(\theta_2-2\psi_2)	= \S_2/\Q2.
\end{equation}
The wall 2 constraint is
\begin{equation}
	R_2^2 = 4\Q2\sin^2\psi_2 + (1-\Q2)^2,
\end{equation}
and for for the time derivative
\begin{equation}
	\d\theta_2 - 2\d\psi_2 = {1\over\S_2}\left(
	\c0\d\theta_0 +  \s0{\d MR_2 - M\d R_2\over R_2^2(1-2M/R_2)}
	\right)
\end{equation}

\subsection{The Effective Action for Two Domain Walls}

Putting all this into our previous action gives
\begin{eqnarray}
&&S=\int dt \biggl\{iR_1^2\d\psi_1 + iR_2^2\d\psi_2 + M\d T\nonumber\\
	&&\qquad\quad-\tilde N^t_1\left[\mu^2 R_1^2 - 4\Q1\sin^2\psi_1 -
		\left(1-\Q1\right)^2\right]	\label{finact}		\\
	&&\qquad\quad-\tilde N^t_2\left[\mu^2 R_2^2 - 4\Q2\sin^2\psi_2 -
		\left(1-\Q2\right)^2\right]\biggr\},\nonumber
\end{eqnarray}
as the effective action for two domain walls, with canonically
conjugate phase space coordinates \footnote{
In this action, the coordinates $\psi_j$ and factors of $i$ have been chosen
appropriate to a tunnelling problem.  The turning points to the classically
allowed regimes of phase space are at
$\psi_j=\pm\pi/2,\pm3\pi/2,\pm5\pi/2\ldots$ and, if desired, the coordinates
can be analytically continued at those points.  See Sect.~IV.}
$(T,i\psi_1,i\psi_2,M,R_1^2,R_2^2)$, in which the gravitational
field is reduced to the single degree of freedom $T$, and in which there is
one degree of freedom $\psi_j$ belonging to each of the two walls.  There
remain two constraints.

We note the following simplifications that have taken place along the
way in the derivation of this action:  1) The original,
non-gauge-invariant, canonical coordinates $(r_j,p_j)$ of the walls
have dropped out completely, in favor of effective, gauge-invariant,
canonical coordinates $(i\psi_j,R_j)$\@.  2) The gravitational degrees
of freedom have all been integrated out, to leave behind just
$(T,M$)\@.  3) The gauge dependent coordinates $i\theta_j$ which arose
during solution of the constraints have likewise disappeared.

\section{Quantization and the Wave Function of the Two Wall System}

Starting from the action given by Eq.~(\ref{finact}), we now proceed
to quantize our 2-wall system, set boundary conditions appropriate
to the initial state of the system, solve for the 2-wall wave function,
$\Psi(\psi_1,\psi_2,T)$ and so determine the decay amplitude.  This
system can be quantized exactly and it is unnecessary to resort to the
WKB approximation.\footnote{But we have indeed made some choices in the
formulation of the constraints, which make no difference to the
classical system, but which do affect details the quantum system
beyond the WKB approximation.  See Sect.~IV for further elaboration.}
The action, Eq.~(\ref{finact}), contains two constraints, which can
be treated by Dirac quantization (See Appendix A).

\subsection{Reduction to $M=0$}

In the initial state --- a single VIS domain wall at minimum radius ---
we have $M$=0\@.  Thus the boundary value of the wave function $\Psi$
is independent of $T$, since $(T,M)$ are canonically conjugate
\cite{KT,KT2,Karel,Don}.
But $M$ commutes with the constraints since $T$ is a cyclic coordinate,
so $\Psi$ must be independent of $T$ throughout:
\begin{equation}
	\Psi(\psi_1,\psi_2,T)=\Psi(\psi_1,\psi_2)
\end{equation}


Thus the domain wall decay problem reduces to a 4-dimensional phase
space in $(\psi_1,\psi_2,R_1,R_2)$ with action
\begin{equation}
S=\int dt \biggl\{iR_1^2\d\psi_1 + iR_2^2\d\psi_2
	-\tilde N^t_1\left[\mu^2 R_1^2 -
	4\sin^2\psi_1 
			\right]
	-\tilde N^t_2\left[\mu^2 R_2^2 -
	4\sin^2\psi_2 
			\right].\biggr\}
\label{mzeroact}
\end{equation}

The canonical coordinates are  now $(i\psi_1,i\psi_2)$ and the canonical
momenta are
\begin{eqnarray}
	&&\pi_1=R_1^2,	\nonumber\\
	&&\pi_2=R_2^2,	
\end{eqnarray}
so that the constraints can be rewritten
\begin{eqnarray}
	&&{\H}_t^1=-i\mu^2\pi_1-4\sin^2(\psi_1)=0,	\nonumber\\
	&&{\H}_t^2=-i\mu^2\pi_2-4\sin^2(\psi_2)=0.
\end{eqnarray}

\subsection{Quantization}

Quantizing by Dirac's procedure, we promote the constraints to operator
equations to define the physical state space,
\begin{eqnarray}
	{\cal H}_t^1\left|\Psi\right\rangle &=& 0,	\nonumber\\
	{\cal H}_t^2\left|\Psi\right\rangle &=& 0.
\end{eqnarray}
We thus take momenta as differential operators
\begin{eqnarray}
	&&\pi_1\to-i{\partial\over{\partial i\psi_1}}=
	-{\partial\over{\partial \psi_1}},		\nonumber\\
	&&\pi_2\to-i{\partial\over{\partial i\psi_2}}.
	-{\partial\over{\partial \psi_2}},
\end{eqnarray}
acting on
the wave function $\Psi(\psi_1,\psi_2)$, and we arrive at the two wave
equations
\begin{eqnarray}
	\left[\mu^2{\partial\over{\partial\psi_1}}
	+ 4\sin^2(\psi_1)\right]\Psi(\psi_1,\psi_2)&=&0,\nonumber\\
	\left[\mu^2{\partial\over{\partial\psi_2}}
	+ 4\sin^2(\psi_2)\right]\Psi(\psi_1,\psi_2)&=&0,
\end{eqnarray}

Since these equations are uncoupled, the wave function separates
as the product of 1-wall wave functions, {\it i.e.,}
\begin{equation}
	\Psi(\psi_1,\psi_2)=\Psi_1(\psi_1)\Psi_2(\psi_2).
\label{waveprod}\end{equation}
It may seem surprising that the two walls are uncoupled from each
other; however this is a necessary consequence of Birkhoff's theorem
and conservation of $M$\@.  We point out two facts:  First, the two
walls are {\it topologically} coupled by residing in the same space;
second, there will be a local effective coupling between the two walls
({\it i.e.,} a $\delta$-function coupling) which will mediate their
annihilation into pure field energy, that we have missed in the thin-wall
approximation.  Then our wave equations become
\begin{equation}
	\left[\mu^2{\partial\over{\partial\psi_j}}
	+ 4\sin^2(\psi_j)\right]\Psi(\psi_j)=0.
\label{waveqs}
\end{equation}

\subsection{Boundary Conditions}

We now set appropriate boundary conditions
for the wave functions whose dynamics are given exactly by
Eq.~(\ref{waveqs})\@.  These equations take the slightly unusual form
of first-order equations, not the usual second-order Schr\"odinger
equation.  First-order equations can, however, be routinely handled by
Dirac quantization \cite{Dirac1}.
\footnote{A slightly different, second-order form of the action
will be presented in Sect.~IV\@.  There is also still another
way to proceed that also leads
to second-order Schr\"odinger equations.  We can put $\pm$
signs on the radicals in Eq.(\ref{finact}), to impose two separate
constraints at each wall instead of one.  Classically,
a solution would be such that one of these constraints vanishes;
the wave function, representing a linear combination of such
allowed states, will only be annihilated by the product of the
constraints.  Hence for the purpose of quantization, we write
\begin{equation}
	{\H}_t^j={\H}_t^{i(+)}{\H}_t^{i(-)}
 		=(\mu R_j)^4-4(\mu R_j)^2+4\sin^2(2\psi_j).
\end{equation}
This eventually leads to equivalent results (up to factor ordering
ambiguities) after careful imposition of boundary conditions.}
It is clear that, in the tunnelling regime, solutions
will die exponentially as $\psi_j$ increases.  The usual freedom
to choose either exponentially dying or exponentially growing
solutions appears in a slightly different way here:  We can choose
a $\psi_j$ in the initial state corresponding to the desired value
of $R_j$, and then evolve either to increasing $\psi_j$ or to
decreasing $\psi_j$, respectively.   Since our problem involves
tunnelling, we wish to evolve both $\psi_j$ to increasing
values.

We now restore the factors of $G$, where
\begin{equation}
	\mu^2G^3\sim\left({m_{\hbox{gut}}
		\over m_{\hbox{pl}}}\right)^6.
\end{equation}
The particular problem we wish to solve involves a pre-existing wall
at the turning point $R_1=2/\mu G$, and a second wall just nucleating at
zero size, $R_2=0$.  Classically from Eq.~(\ref{mzeroact}),
\begin{equation}
	R_j={2\over\mu G}\sin\psi_j,
\end{equation}
so translating the $R_j$ to canonical wall coordinates,
\begin{eqnarray}
	\psi_1 &=& \pi/2,	\nonumber\\
	\psi_2 &=& 0,
\end{eqnarray}
which give the initial conditions on the wave function.
\begin{equation}
	\Psi(\pi/2,0) = c_{\rm nuc},
\label{initcond}\end{equation}
where $c_{\rm nuc}$ is the amplitude to nucleate a zero-size
wall, which we take to be $\sim1$\@.

The tunnelling process then proceeds with both $\psi_j$ increasing,
until the walls meet and annihilate.  This necessitates for the
final state
\begin{equation}
	R_1 = R_2,
\end{equation}
or (taking into account the multiple-valuedness of $\sin$)
\begin{equation}
	\psi_1 = \left\{\begin{array}{l}\psi_2\\
				    \pi-\psi_2\end{array}\right.
\end{equation}
The allowed domain for the canonical coordinates is
\begin{eqnarray}
	\pi/2	&\le& \psi_1 \le \pi,	\nonumber\\
	0	&\le& \psi_1 \le \pi/2
\label{domain}\end{eqnarray}
so the final condition must be
\begin{equation}
	\psi_1 = \pi-\psi_2.
\label{finalcond}\end{equation}
The final state is not a turning point of the classical action,
Eq.~(\ref{mzeroact}), because the annihilation process is missed
in this action, as mentioned above.  Presumably a small-scale
turning point would appear if we worked beyond the thin-wall
approximation, but we will not pursue this point.  Figure 2
displays the configuration space of the 2-wall system in the
tunnelling regime.
\bfig{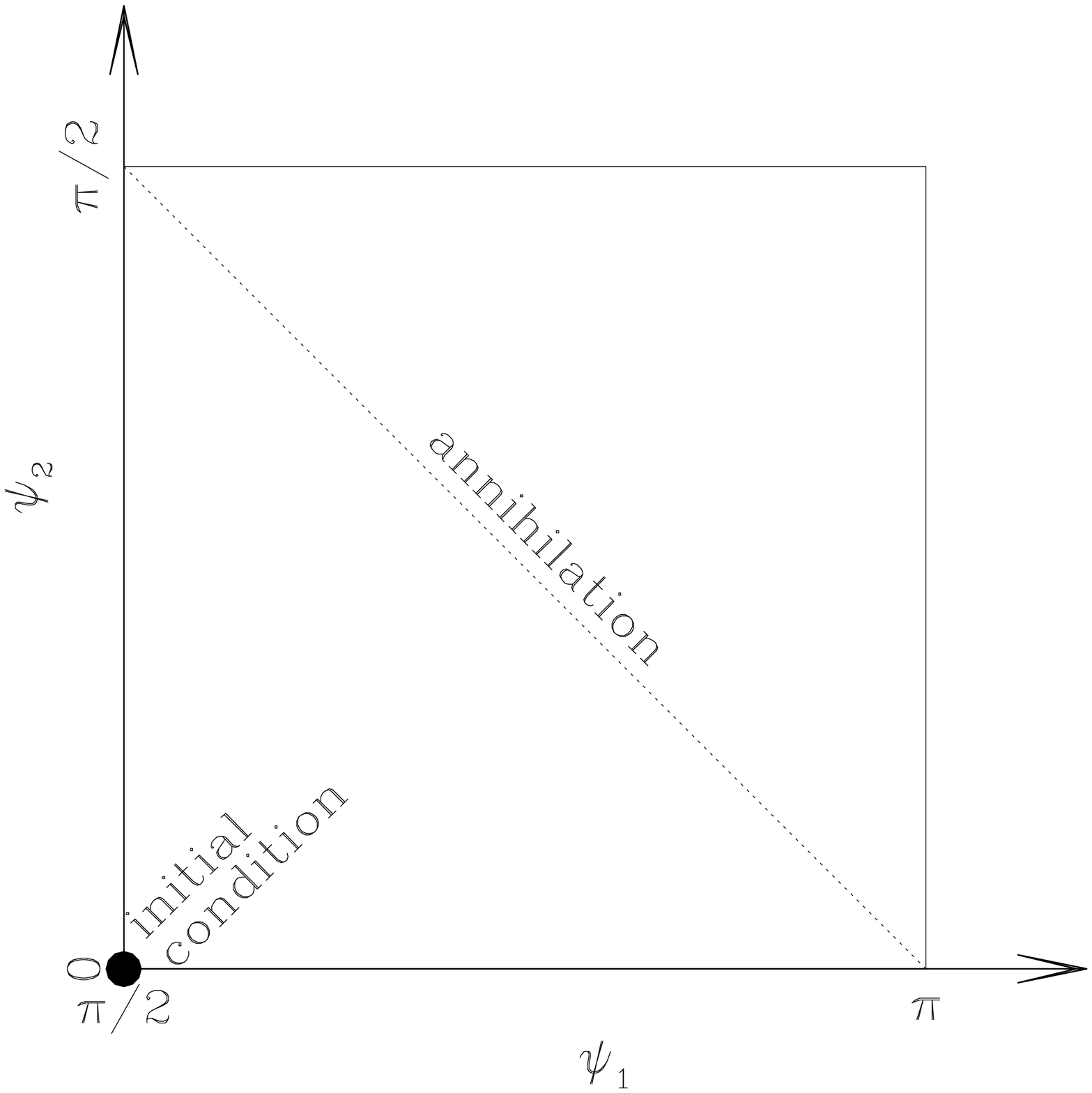,width=5.5in}
\vskip 1cm
\caption{The configuration space of the two wall system in the
tunnelling regime.  The coordinates are $\psi_1$ and $\psi_2\,$.
Initial conditions are shown as $\bullet$ at the lower left.  The
diagonal  dotted line represents the locus where the walls collide
and annihiliate into pure field energy.}
\label{q2fig2}
\efig

\subsection{Solution of the Wave Equations}

The wave equations (\ref{waveprod},\ref{waveqs}) comprise two independent
first-order equations in the $(\psi_1,\psi_2)$ plane, which are well
posed and have a unique solution under the the initial conditions
(\ref{initcond}) on the domain (\ref{domain}).
This solution, for the wave function of the 2-wall system, is then
\begin{equation}
  \Psi(\psi_1,\psi_2) = c_{\rm nuc}\exp\left(-{1\over\mu^2G^3}\biggl[
	2\psi_1 + 2\psi_2 -\pi -\sin2\psi_1 - \sin2\psi_2\biggr]\right)
\label{wavesoln}
\end{equation}
Evaluating this wave function under the final condition,
Eq.~(\ref{finalcond}) gives
\begin{equation}
	\Psi_{\rm final}(\psi_1,\pi-\psi_1) =
		c_{\rm nuc}\exp\left({-\pi/\mu^2G^3}\right)
\label{wavefinal}
\end{equation}
for the final state wave function.  The corresponding probability
of tunnelling to the final state is
\begin{equation}
	P_{\rm final} = \left|c_{\rm nuc}\right|^2
			\exp\left({-2\pi/\mu^2G^3}\right)
\label{probfinal}
\end{equation}
The most remarkable feature of the result of this paper,
Eq.~(\ref{probfinal}), is its
{\it independence} of the final value of $\psi_1$ or $\psi_2$\@.
This means that the two walls may collide and annihilate at any value
of the final radius in the kinematically allowed range,
\begin{equation}
	0\le R_{\rm final}\le2/\mu G,
\label{radfinal}
\end{equation}
with equal probability.  At first this may seem surprising, but we
argue that it is as expected.

Consider the following toy problem.  A particle and an antiparticle
move in a potential that is identical for both particles. (For instance,
a proton and an antiproton move in a gravitational potential.)  There
is a potential barrier present, and in the initial state, the two
particles are on opposite sides of this barrier.  They may tunnel
toward each other through the barrier, and annihilate if they meet.
The question now is, what is the most probable location for the
annihilation?  We encourge the reader to stop reading at this point,
guess the answer, and then work it out.

The answer is that annihilation is equally probable at any location
within the barrier, and the annihilation probability is just given
by the total barrier factor for single-particle penetration.  We argue
that annihilation of the 2-wall system is no different, justifying
our result.  However, this system is not easy to interpret
in the canonical variables $(i\psi,R^2)$\@.  Therefore we will
also give a slightly different quantization for the 1-all system, with
application to the 2-wall system, after defining some new
canonical variables.

\section{A Further Method of Quantization}

A better pair of variables $(Q,P)$ for the 1-walll system can be obtained
by defining
\begin{equation}
	\chi = i(\pi/2-\psi)
\end{equation}
and then carring out the following canonical transformation:
\begin{eqnarray}
	Q &=&  \sqrt2 R \cosh\chi,				\nonumber\\
	P &=&  \sqrt2 R \sinh\chi,
\label{defQP}
\end{eqnarray}
which entail
\begin{eqnarray}
  iR^2\dot\psi	&=& P\dot Q + \hbox{(total time derivative),}	\nonumber\\
 \mu^2R^4 - 4R^2\sin^2\psi &=&
	 \mu^2R^4 - 4R^2\cosh^2\chi,				\nonumber\\
	&=& {\mu^2\over4}(Q^2-P^2)^2 - 2Q^2,			\nonumber\\
	&=& \left({1\over2}(Q^2-P^2) + {\sqrt2\over\mu} Q\right),
	    \left({1\over2}(Q^2-P^2) - {\sqrt2\over\mu} Q\right).
\label{transQP}
\end{eqnarray}
(We are again setting $G=1$\@.)  The 1-wall action then can be rewritten
as
\begin{eqnarray}
      S	&=& \int dt \left\{
	iR^2\dot\psi - \tilde N^t\left[\mu^2 R^2 - 4\Q1\sin^2\psi -
		\left(1-\Q1\right)^2\right]\right\}		\nonumber\\
	&=& \int dt \left\{P\dot Q - \bar N^t \left({1\over2}(Q^2-P^2) -
		{\sqrt2\over\mu} Q\right)\right\}
	\equiv \int dt \left\{P\dot Q - \bar N^t \bar{\cal H}_t\right\},
\label{PQact}
\end{eqnarray}
where the constraint is redefined by
\begin{eqnarray}
  \tilde {\cal H}_t = \bar {\cal H}_tR^{-2}
	\left({1\over2}(Q^2-P^2) + {\sqrt2\over\mu} Q\right),	\nonumber\\
  \bar N^t = \tilde N^t R^{-2}
	\left({1\over2}(Q^2-P^2) + {\sqrt2\over\mu} Q\right).
\label{barH}
\end{eqnarray}
Once more, the extra factors introduced into the constraint by
Eq.~(\ref{barH}) never vanish in the classical regime, and make no
difference to the classical equations of motion; and they never vanish
in the quantum tunnelling regime and so at most affect the prefactor in
the tunnelling calculation.
The phase space $(Q,P)$ of the 1-wall action in the form (\ref{PQact})
can be described as follows.  The classically allowed regime is
\begin{equation}
	Q \hbox{ real, } 0\le Q<\infty; \qquad
	P \hbox{ real, } 0\le\left|P\right|<Q;
\end{equation}
while the classically forbidden, or quantum tunnelling, regime is
\begin{equation}
	Q \hbox{ real, } -\infty\le Q<\infty;\qquad
	P \hbox{ imaginary, } -i\infty\le P<i\infty.
\end{equation}
The two regimes meet at $P=0$.

The action (\ref{PQact}) is now entirely straightforward to quantize
as a 1-dimensional particle system.  We take the 1-wall wave function
as $\Psi(Q)$ and use
\begin{equation}
	P \to-i {\partial\over\partial Q}
\end{equation}
to write the constraint as a Schr\"odinger equation ${\cal H}_t\Psi=0$ or
\begin{equation}
 \left( - {\partial^2\over\partial Q^2} + V(Q)\right)\Psi(Q)=0,
\label{PQwaveqn}
\end{equation}
where the potential is an ``upside-down harmonic oscillator".
\begin{eqnarray}
  V(Q)	&=& - Q^2 + {2\sqrt2\over\mu}Q,			\nonumber\\
	&=& -\left(Q-{\sqrt2\over\mu}\right)^2 + {2\over\mu^2}.
\end{eqnarray}
Note that the constraint says that $\Psi$ must be the zero eigenfunction
of ${\cal H}_t$; the rest of the spectrum of ${\cal H}_t$ is not present.
Equation (\ref{PQwaveqn}) is hypergeometric and its solutions are
parabolic cylinder functions, but we will not pursue the details.
Figure 3 shows the potential $V(Q)$ and the resultant dynamics.
\bfig{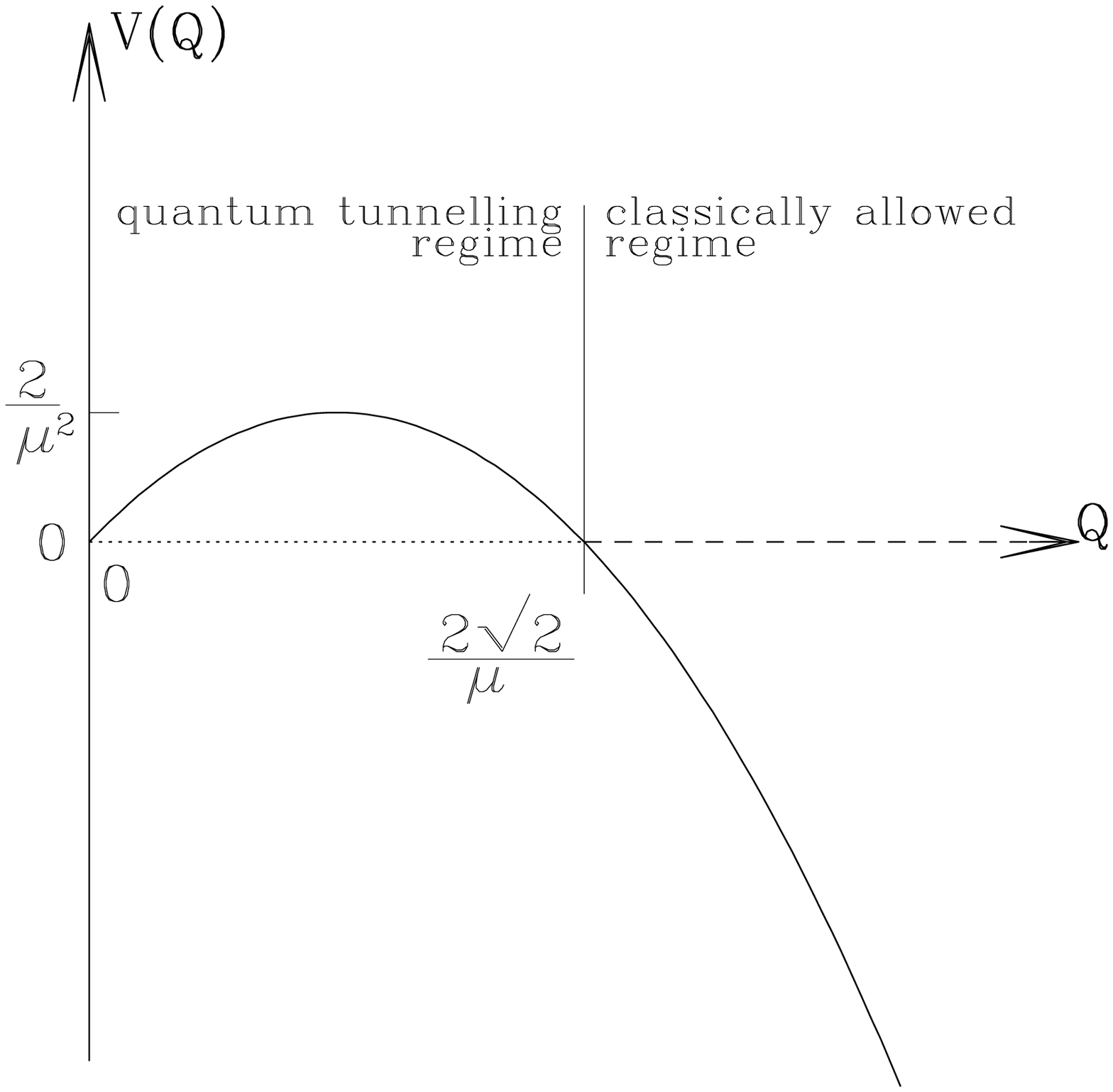,width=5.5in}
\vskip 1cm
\caption[]{Dynamics of the quantized VIS solution.
The domain wall can be viewed as a particle moving in one dimension
$Q$, under the influence of a potential $V(Q)$ (curve).  The energy
is constrained to be 0\@.  The turning point is at
$Q=2\sqrt2/\mu$;  to its right is the classically allowed
regime, and to its left is the  classically forbidden, or quantum
tunnelling, regime.  The dashed horizontal line represents the
classical motion of the VIS wall, and the dotted line represents
quantum tunnelling.}
\label{q2fig3}
\efig

Tunnelling from $Q=0$ to $Q=2\sqrt2/\mu$ is the ``creation of the
VIS universe from nothing", while tunnelling in the opposite direction
is ``annihilation of the VIS universe into nothing"
({\it cf.} \cite{andrew}).

To study the quantum decay of the VIS universe into a universe containing
pure field energy, our subject in this paper, two copies of the 1-wall
system must be coupled to make the 2-wall system with configuration
space $(Q_1,Q_2)$.  Boundary conditions are as follows:  For wall 1,
impose along the $Q_1$-axis purely right-going boundary conditions on
the left.  For wall 2, impose along the $Q_2$-axis purely left-going
boundary conditions on the right.  Annihilation can occur at any $Q_1=Q_2$
between the endpoints $0$ and $2\sqrt2/\mu$\@.  The WKB approximation gives
the same exponential barrier factor
$P\sim\exp(-\pi/\mu^2G^3)$ as in Eq.~(\ref{probfinal}) above, and
the same main result appears, that the decay probability is independent
of wall radius at annihilation (or final $Q$)\@.  We will not pursue
the details beyond the WKB regime; presumably the prefactor will differ.

\section{Conclusion}

The main results of this paper are:
\begin{enumerate}
  \item	The 1-wall VIS universe does undergo quantum decay into a universe
	containing pure field energy, with some small probability.
  \item	The decay process can be treated as the nucleation of a second
	domain wall at zero size, followed by quantum tunnelling of the
	two walls toward each other, and anniliation when they meet.
  \item	The 2-wall system can be treated in a Hamiltonian approach,
	using a simple action in the 2-wall phase space, and Dirac
	quantization.
  \item The decay probability for the VIS universe is independent of
	the radius of the final universe (up to prefactors), and is given
	by Equation (\ref{probfinal}).
\end{enumerate}

Conclusions 1\ and 2\ agree with paper I \cite{paperI}, which employs
an instanton approach to replace 3\@.  However, conclusion 4\
differs strongly from paper I, which predicted a unique value
\begin{equation}
	R_{\rm final}=\sqrt2/\mu G
\end{equation}
for the annihilation radius.
Furthermore, the probability differs:  The above value,
Eq.~(\ref{probfinal}) differs from the value of paper I, Eq.~(I.88), which is
\begin{equation}
  P_{\hbox{paper I}}
	\sim \exp\left(-{2\over{\mu^2 G^3}}\right).
\end{equation}
We interpret this disagreement as an incorrect result of the instanton
approach to this problem (at least as done in paper I)\@.

In fact, the two results
can be reconciled, if we imitate the Hamiltonian calculation and flip
some signs in the instanton calculation, in a way that seems
{\it ad hoc} in the instanton context.  In particular, if the four
segments in the $n=2$ instanton are weighted $(+1,-1,+1,-1)$ in
calculating the action (see paper I), rather than using the
$(+1,+1,+1,+1)$ that was previously motivated both by the standard
methodology of Euclidean quantum gravity, and by the 2-sheeted
manifold rule of Farhi, Guth and Guven \cite{FGG}.  These sign flips
are now motivated by a careful consideration of exponentially
growing versus exponentially dying wave functions.  After the
signs are flipped, the instanton decay probability agrees with
Eq.(\ref{probfinal})\@.  We conclude
that the instanton method as utilized in paper I makes incorrect
choices for these wave functions.  We would like to be able to
propose a ``modified rule" for sign weights in the instanton
calculation
that would repair this defect, but have been unable to find a
convincing formuation.

We leave the correct instanton treatment of the quantum decay of
domain walls as an open problem.

Our result (\ref{probfinal}) also disagrees with the answer one obtains
by assuming that the tunnelling probability is of the form
\begin{equation}
	P\sim\exp(I_f-I_i),
\end{equation}
where $I_f$ and $I_i$ are the Euclidean actions of the instantons which
mediate the creation from nothing of the final state and the initial
state, respectively, as is often done.  But, our answer does happen to
be the same as the probability $\exp(-I_j)$ for creation of the initial
state alone from nothing, as calculated by using the Euclidean VIS
solution as the instanton \cite{andrew}.  It is not clear why the various
methods do not agree.

The barrier factor we have calculated is a function
of the dimensionless parameter
$\mu^{-2}G^{-3}\sim(m_{\hbox{gut}}/m_{\hbox{pl}})^6$.
For a typical theory, $m_{\hbox{gut}}\sim 10^{14}\hbox{---}10^{18}$ GeV;
hence $\mu^{-2}G^{-3}$
is expected to be extremely small in most phenomenologically viable
models of microphysics.  However, improbable events can be
important in early-universe cosmology, if they lead to a universe
resembling our own.

The new universe created by the decay does not yet resemble our
own, however.  To do so it must first expand greatly, and then
it must homogenize itself.  Whether it does so will be the subject
of a future paper in this series.

\acknowledgements

This research was supported in part by the National Science
Foundation under Grant Nos.~PHY89-04035 and PHY90-08502\@.
We are grateful to Joe Polchinski, Matthew Fisher and Andrew Chamblin
for helpful conversations.  DME is grateful to the Aspen Center for
Physics where part of the work was carried out.

\appendix
\section{Quantization of Gauge Theories}


Dirac first worked out the theory of quantizing constrained systems
in general \cite{Dirac1}, and general relativity in particular \cite{Dirac2},
and his pioneering work continues to serve as the foundation of current
efforts to canonically quantize gravity.  What follows will be a very
brief review of the main elements of such a quantization scheme,
sufficient for the purposes of the current study.  Many more extensive
studies of the subject can be found in the literature; see, for
example, \cite{Dirac3,HRT,Sundermeyer}.

Consider a mechanical system with n degrees of freedom, which is
described by the Lagrangian $L( q_i,\dot q_i)$, and
where the $q_i(t),\; i=1,\dots,n$ are generalized coordinates.
The canonical momenta are defined by
\begin{equation}
	 p_i = {{\partial L}\over{\partial\dot q_i}}( q_i,\dot q_i).
\label{momeqs}
\end{equation}
To put this action into Hamiltonian form, one seeks to eliminate the
velocities $\dot q_i$ in favor of the momenta $ p_i$ through the
use of Eqs.~(\ref{momeqs}).  However, in the event that the Hessian
matrix of $L$ with respect to the velocities has zero determinant,
{\it i.e.,}
\begin{equation}
	\hbox{Det} {\H}_{ij}= \hbox{Det}{{\partial^2 L}\over
		{\partial\dot q_i\partial\dot q_j}}=0,
\end{equation}
then not all of the $\dot q_i$ can be eliminated in this manner.
(This will occur, for example, if the action is linear in one or
more of the velocities, and quadratic in the rest.)  In fact, one
can eliminate exactly $R$ of the $\dot q_i$, where $R<n$ is the
rank of ${\H}_{ij}$.  After doing so, one is left with a set of
$(n-R)$ constraint equations of the form
\begin{equation}
	C_\alpha\equiv p_\alpha - f_\alpha( q_i, p_i) = 0,
		\; \alpha=1,\dots,n-R,
\end{equation}
which are known as the {\it primary} constraints of the theory.

The canonical Hamiltonian,
\begin{equation}
	H_c\equiv \sum_{i=1}^R p_i\dot q_i - L
\end{equation}
is not unique on the full phase space $(q_i,p_i)$, and so one
defines the new Hamiltonian
\begin{equation}
	H = H_c +\lambda^\alpha C_\alpha,
\end{equation}
where the $\lambda^\alpha$ are arbitrary functions, or Lagrange
multipliers.

Introduce the notation $\{u,v\}$ as the Poisson bracket of the functions
$u(p,q)$ and $v(p,q)$, and let $\{u,v\}'$ denote a Poisson bracket to which
the constraints have been applied after the calculation of the bracket.
Then one divides the primary constraints into two classes, according
to the algebra of their Poisson brackets.  Those constraints whose
Poisson bracket algebra closes, {\it i.e.,} for which
\begin{equation}
	\{C_\alpha,C_\beta\}=f_{\alpha\beta}^\gamma C_\gamma,
\end{equation}
or
\begin{equation}
	\{C_\alpha,C_\beta\}'=0,
\end{equation}
are known as {\it first-class} constraints, and all other constraints are
known as {\it second-class}.  We will denote a second-class constraint with
a Latin index, {\it e.g.,} $C_a$.  An important fact is that the for each
first-class constraint there is a corresponding gauge symmetry of the
theory.

Since the constraints should hold at
all times, we require that
\begin{equation}
	\dot C_\alpha=\{C_\alpha,H\}' = 0,
\label{Cdot}
\end{equation}
which may in some cases lead to inconsistent equations, or which may
lead to new relations between the phase space variables.  In the latter
case, these new relations are known as {\it secondary} constraints,
which are then classified as first-class or second-class as described
above.  The consistency conditions (\ref{Cdot}) must then be checked
again, and the process repeated until all constraints have been found.

In order to quantize the theory, we would like to replace the
Poisson bracket in the classical relations by $-i/\hbar$ times the
commutator of the corresponding quantum operators, and then impose the
constraints as conditions on the state vectors.  However, note that
\begin{eqnarray}
	&&C_\alpha|\psi\rangle=0, \;\; C_\beta|\psi\rangle=0\nonumber\\
	&&\Rightarrow [C_\alpha,C_\beta]|\psi\rangle=0,
\end{eqnarray}
which corresponds to the classical relation
\begin{equation}
	\{C_\alpha,C_\beta\}'=0.
\end{equation}
Hence all of the constraints should be first-class in order for the
quantization to go through in a straightforward way.

The prescription for eliminating the second-class constraints is as follows.
The Dirac bracket is defined by
\begin{equation}
	\{A,B\}^*=\{A,B\} - \{A,C_a\}\Gamma^{ab}
		\{C_b,B\}
\end{equation}
where the algebra of the second-class constraints is
\begin{equation}
	\Gamma^{ab}\{C_b,C_c\}=\delta^a_c.
\end{equation}
This is a projection of the Poisson bracket onto the second-class
constraint surface; therefore if the Poisson brackets are replaced with
Dirac brackets in the classical analysis, we can consistently take
$C_a=0$ to hold as operator equations in the quantum theory.  Finally
the quantization can proceed, now with the Dirac bracket taking the
role of the Poisson bracket in the classical theory:
\begin{eqnarray}
	&&\{u,v\}^*\to -{i\over\hbar}[u,v],	\nonumber\\
	&&p_i\to{\hbar\over i}{\partial\over{\partial q_i}},\\
	&&C_\alpha(p_i,q_i)|\psi\rangle=0.		\nonumber
\end{eqnarray}


\end{document}